\documentclass[fleqn,usenatbib]{mnras}

\usepackage{newtxtext,newtxmath}
\usepackage[T1]{fontenc}

\DeclareRobustCommand{\VAN}[3]{#2}
\let\VANthebibliography\thebibliography
\def\thebibliography{\DeclareRobustCommand{\VAN}[3]{##3}\VANthebibliography}

\usepackage{graphicx}	% Including figure files
\usepackage{amsmath}	% Advanced maths commands

\def\deg{\ifmmode^\circ\else$^\circ$\fi}
\def\kms{km\thinspace s$^{-1}$}

\def\Q{\ifmmode\mathcal{Q}\else$\mathcal{Q}$\fi}
\def\Mach{\ifmmode\mathcal{M}\else$\mathcal{M}$\fi}

\title[A dual dense probably rotating outflow from a MYSO]
{ALMA discovery of a dual dense probably rotating outflow from a massive young stellar object G18.88MME}

\author[Zinchenko et al.]{
I. I. Zinchenko$^{1}$\thanks{E-mail: zin@iapras.ru},
L.~K. Dewangan$^{2}$, T. Baug$^{3}$, D.~K. Ojha$^{4}$, and N.~K. Bhadari$^{2,5}$\\
$^1$Institute of Applied Physics of the Russian Academy of Sciences 46 Ul'yanov~str., Nizhny Novgorod 603950, Russia.\\
$^2$Physical Research Laboratory, Navrangpura, Ahmedabad - 380 009, India.\\
$^3$Satyendra Nath Bose National Centre for Basic Sciences, Block-JD, Sector-III, Salt Lake, Kolkata-700 106, India.\\
$^4$Department of Astronomy and Astrophysics, Tata Institute of Fundamental Research, Homi Bhabha Road, 
Mumbai 400 005, India.\\
$^5$Indian Institute of Technology Gandhinagar Palaj, Gandhinagar, 382355, India.\\
}

\date{Accepted XXX. Received YYY; in original form ZZZ}

\pubyear{2021}

\begin{document}
\label{firstpage}
\pagerange{\pageref{firstpage}--\pageref{lastpage}}
\maketitle

\begin{abstract}
We report the discovery of a very dense jet-like fast molecular outflow surrounded by a 
wide-angle wind in a massive young stellar object (MYSO) G18.88MME ({stellar} mass $\sim$8 M$_{\odot}$) powering 
an Extended Green Object G18.89$-$0.47. 
{Four cores MM1--4} are identified in the Atacama Large Millimeter/submillimeter Array (ALMA) 1.3 mm continuum map 
(resolution $\sim$0\rlap.{$''$}8) toward G18.88MME, and are seen at {the center of the emission structure ({extent $\sim$0.3 pc $\times$ 0.2 pc}) detected in the ALMA map}.
G18.88MME is embedded in the core MM1 (mass $\sim${13--18} M$_{\odot}$), where no radio continuum emission is detected. The molecular outflow centered at MM1 is investigated in the SiO(5--4), HC$_{3}$N(24--23) and $^{13}$CO(2--1) lines. 
The detection of HC$_{3}$N in the outflow is {rare} in MYSOs and indicates its very high density. 
{The position-velocity diagrams} display a fast narrow outflow (extent $\sim$28000 AU) and a slower wide-angle {more extended} outflow toward MM1, and both of these components {show a transverse velocity gradient indicative of a possible rotation.}
%appear to be rotating. 
All these observed features together make 
G18.88MME as a unique object for studying the unification of the jet-driven and wind-driven scenarios of molecular outflows in MYSOs.
\end{abstract}


%%%%%%%%%%%%%%%%%%%%%%%%%%%%%%%%%%%%%%%%%%%%%%%%%%

%%%%%%%%%%%%%%%%% BODY OF PAPER %%%%%%%%%%%%%%%%%%

\section{Introduction}
\label{sec:intro}
Our understanding of the physical processes involved in the formation of massive stars ($>$ 8 M$_{\odot}$) is still incomplete \citep[e.g.,][]{zinnecker07,Motte+2018,rosen20}. Most recent high-resolution studies of massive young stellar objects (MYSOs) suggest that 
massive stars form via infall from a surrounding envelope and disk-mediated accretion similar to their low-mass counterparts \citep[see][]{rosen20}. 
In this relation, understanding accretion and jets/outflows in MYSOs is crucial for constraining the physics related to their birth 
and early evolution. 
The simultaneous detection of a highly collimated jet and a wide-angle outflow in low-mass stars has 
been known \citep[e.g.,][]{arce07,zapata14} and theoretical models of such event have been developed. 
These are considered as two flavours of the same mass-ejection process in protostars. 
However, the study of connection between these two forms in MYSOs {is still lacking \citep[e.g.,][]{arce07,frank14} and} {very few observations of the simultaneous presence of a highly collimated jet and a wide-angle outflow in MYSOs exist \citep[e.g.,][]{torrelles11,zinchenko20}.}
%such event is also extremely rare in MYSOs \citep[e.g.,][]{torrelles11}. 
No models predict the simultaneous existence of a jet and a wide-angle outflow as well as rotation of both the components together in MYSOs \citep[e.g.,][]{torrelles11}. Therefore, the observational study of such two flavours of ejection from MYSOs can provide vital inputs for constraining the accretion based models for massive star formation. 
{In this context, the target of this paper is a genuine MYSO embedded in the infrared dark cloud (IRDC) SDC18.888--0.476 (hereafter SDC18). The target MYSO is associated with both the 6.7 GHz methanol maser emission (MME) and Extended Green Object (EGO).} 

Situated at a distance of $\sim$5.0 kpc, the IRDC SDC18 has been seen as a bright emission region in the submillimeter (sub-mm) maps \citep[see Figure~2 in][]{dewangan20c}. 
A massive dust clump at 870 $\mu$m has been reported toward SDC18 \citep[$L_\mathrm{bol}$ $\sim$53.8 $\times$ 10$^{3}$ $L_\odot$; $M_\mathrm{clump}$ $\sim$4880 $M_\odot$;][]{urquhart18,dewangan20c}, 
which is located at the edge of the W39 H\,{\sc ii} region hosting several massive OB-stars \citep{westerhout58,kerton13}. Figure~\ref{fig1}a displays the {\it Spitzer} 24 $\mu$m image overlaid with the IRAM 1.2 mm continuum emission contours \citep[beam size $\sim$13$''$;][]{rigby18}, showing the mm emission peaks toward the clump at 870 $\mu$m. 
In Figure~\ref{fig1}a, we mark the positions of a water maser \citep[V$_\mathrm{lsr}$ $\sim$65.1 km s$^{-1}$;][]{walsh11} and a 6.7 GHz MME \citep[V$_\mathrm{lsr}$ $\sim$56.4 km s$^{-1}$;][]{breen15,yang19}.
The position of the 6.7 GHz MME coincides with {the peak of 1.2 mm continuum emission} and the EGO G18.89$-$0.47 \citep{cyganowski08,towner19}. 
{\citet{dewangan20c} identified a point-like source as an infrared counterpart (IRc) of the 6.7 GHz MME (hereafter G18.88MME). Previously, \citet{kerton13} characterized this source as a protostar (stellar mass $\sim$8 M$_{\odot}$; see source ID \#G18-2 in Table~3 in their paper) using the photometric data at 3.6--24 $\mu$m.} 
%
%This source was previously characterized as a protostar (
%Using the photometric data at 3.6--24 $\mu$m, this source was previously characterized 
%\citet{kerton13} identified a protostar (stellar mass $\sim$8 M$_{\odot}$; see source ID \#G18-2 in Table~3 in their paper).  been investigated toward the position of the 6.7 GHz MME \citep[see also][]{dewangan20c}, 
%which was treated as an infrared counterpart (IRc) of the 6.7 GHz MME \citep[hereafter G18.88MME;][]{dewangan20c}.
% 
No radio counterpart of G18.88MME is reported in the literature \citep[e.g.,][]{towner19,dewangan20c}. 
All these results allowed \citet{dewangan20c} to propose the IRc G18.88MME as a genuine MYSO candidate in a very early evolutionary phase, just before the ultracompact (UC) H\,{\sc ii} stage (see an arrow in Figure~\ref{fig1}a). 

In this letter, we analyzed the Atacama Large Millimeter/submillimeter Array (ALMA) 1.38 {mm} continuum map and data cubes of SiO, HC$_{3}$N, and $^{13}$CO lines (resolution $\sim$0\rlap.{$''$}8 or 4000 AU) of G18.88MME, which have enabled us to discover 
%the simultaneous presence of a wide-angle outflow and a collimated jet in the core hosting the exciting source G18.88MME. 
a dense fast narrow jet-like molecular outflow surrounded by a slower wide wind. 
Both components are found to be rotating. 

{The ALMA data adopted in this work are described in Section~\ref{sec:obser}. 
The results derived using the continuum and line data are presented in Section~\ref{sec:results}. 
We discuss the implications of our observed findings in Section~\ref{sec:discussion}.}
%{sec:disc}. 
%In Section~\ref{sec:conc}, we summarize the main results of our study.

%Furthermore, we have also investigated the wide-angle outflow in the HC$_{3}$N emission, which is itself the rarest event. 
%
\section{Data sets}
\label{sec:obser}
We downloaded the ALMA continuum map at 1.38 mm (resolution $\sim$0\rlap.{$''$}82 $\times$ 0\rlap.{$''$}6; P.A. = 66$\degr$.6) 
and the processed cubes of three lines (i.e., SiO(5--4), HC$_{3}$N(24--23), and $^{13}$CO(2--1); beam size $\sim$0\rlap.{$''$}9 $\times$ 0\rlap.{$''$}66) from the ALMA science archive. 
These observations were taken with the 12-m ALMA array in Band-6 under the 
project 2017.1.00983.S (PI: Brogan, Crystal). All these data sets were produced by the ALMA pipeline. 
The SiO(5--4), HC$_{3}$N(24--23), and $^{13}$CO(2--1) lines were observed 
in spectral windows with frequency ranges (bandwidths) of 216.58--217.52 GHz (935 MHz), 217.55--218.49 GHz (468 MHz), and 220.24--220.71 GHz (468 MHz), respectively.
The velocity resolutions (rms noise) of the SiO(5--4), HC$_{3}$N(24--23), and $^{13}$CO(2--1) lines 
{are 1.35 km s$^{-1}$ (2.1 mJy beam$^{-1}$), 0.33 km s$^{-1}$ (2.1 mJy beam$^{-1}$), and 0.33 km s$^{-1}$ (4.1 mJy beam$^{-1}$), respectively.
Additionally, we used the {\it Spitzer} MIPS Inner Galactic Plane Survey \citep[MIPSGAL;][]{carey05} 24 $\mu$m image (resolution $\sim$6$''$) and the IRAM 1.2 mm continuum emission map \citep[resolution $\sim$13$''$; from][]{rigby18} of SDC18.}
\section{Results}
\label{sec:results}
\subsection{Continuum Emission}
\label{sec:conta}
\begin{figure}
\center
\includegraphics[width=7.8cm]{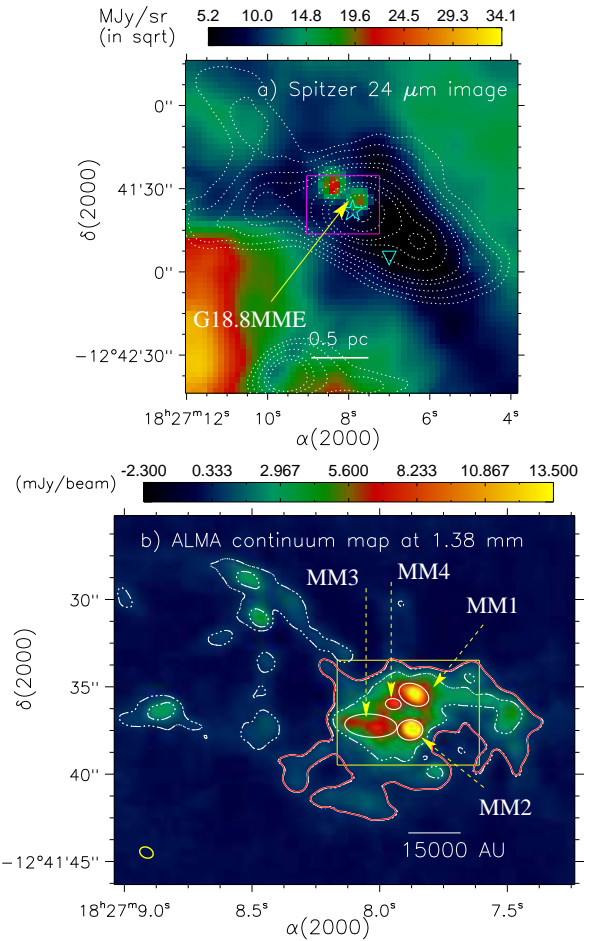}
\caption{a) Overlay of the IRAM 1.2 mm continuum contours on the {\it Spitzer} 24 $\mu$m image of SDC18. 
The 1.2 mm continuum map is exposed to a Gaussian smoothing function with a width of 3 pixels. 
{The contour levels of the IRAM continuum emission are 28, 45, 70, 100, 120, 180, 250, 300, 430, and 570 mJy beam$^{-1}$.} The positions of a water maser and a 6.7 GHz MME are marked by an upside down triangle and a star, respectively. 
The solid box (in magenta) outlines the area shown in Figure~\ref{fig1}b. 
b) The panel shows the ALMA 1.38 mm continuum image around G18.88MME. 
The 1.38 mm continuum contours are also shown with the levels of 0.55 and 2 mJy beam$^{-1}$. 
The synthesized beam is $\sim$0\rlap.{$''$}82 $\times$ 0\rlap.{$''$}6, P.A. = 66$\degr$.6 (lower left corner). 
{{The 1.38 mm continuum contour with a level of 0.55 mJy beam$^{-1}$ (in red) outlines the emission structure.} {Four} continuum sources are indicated by ellipses (in white; see also broken arrows). 
The solid box (in yellow) outlines the area shown in Figures~\ref{fig3}a and~\ref{fig3}b.}} 
\label{fig1}
\end{figure}

To further study the inner structures of the clump {detected} in the IRAM 1.2 mm continuum map (see Figure~\ref{fig1}a), we present the ALMA 1.38 mm continuum map and contours in Figure~\ref{fig1}b. 
{In the direction of G18.88MME, four continuum sources (i.e., MM1, MM2, MM3, and MM4) are identified (see ellipses), and are spatially found at the center of 
the emission structure ({extent $\sim$0.3 pc $\times$ 0.2 pc}; see {the red contour} in Figure~\ref{fig1}b) traced in the ALMA map.} 
%a shell-like feature (extent $\sim$0.36 pc $\times$ 0.19 pc). 
The continuum sources were selected using the Python-based {\sc astrodendro}-package\footnote{https://dendrograms.readthedocs.io/en/stable/index.html} 
\citep[see also][for more details]{baug20}, 
which uses the {\sc dendrogram} algorithm as described in \citet{rosolowsky08}. 
In this context, a minimum flux level of 5$\sigma$ was adopted, where the background flux (i.e., 1$\sigma$) of 0.06 mJy beam$^{-1}$ was estimated from the emission-free areas in the continuum map. 
%In the calculations, we only considered structures extended over an area of more than the {beam area of 33 pixels.} 
%Note that the continuum source MM1 contains another small emission peak. {These peaks are not identified as separate sources because these are located within a one beam size.} % of the continuum map.} 
%., but the extent of its associated continuum source is smaller than the beam size of the continuum map.}

{The flux densities (deconvolved angular sizes) of the continuum sources MM1, MM2, MM3, and MM4 are 
about 33.5 mJy (0\rlap.{$''$}92 $\times$ 0\rlap.{$''$}6), 37.3 mJy (0\rlap.{$''$}75 $\times$ 0\rlap.{$''$}58), 
24.4 mJy (1\rlap.{$''$}53 $\times$ 0\rlap.{$''$}66), and 6.3 mJy (0\rlap.{$''$}46 $\times$ 0\rlap.{$''$}31), respectively.} \citet{towner19} reported the dust temperature ($T_{\rm d}$) and the kinetic 
temperature of the clump hosting G18.88MME to be $\sim$22~K and $\sim$28~K, respectively. The kinetic temperature was derived from the NH$_3$ emission \citep{towner19}. 
The observed mm fluxes ($S_\nu$) can be utilized to compute the masses of continuum sources \citep[e.g.,][]{hildebrand83}. 
In the calculations, we adopted the dust opacity ($\kappa_\nu$) = 0.9\,cm$^2$\,g$^{-1}$ at 1.3 mm \citep{ossenkopf94}, distance ($d$) = 5.0 kpc, and $T_{\rm d}$ = [22, 28]~K \citep[see equation~1 in][]{dewangan16}. 
Using the values of $S_\nu$ and $T_{\rm d}$ = 22(28)~K, {the masses of MM1, MM2, MM3, and MM4 are estimated to be $\sim$18(13), $\sim$20(15),  $\sim$13(10), and $\sim$3.4(2.5) M$_{\odot}$, respectively.} The uncertainty in the mass calculation could be typically $\sim$20\% to $\sim$50\%, which includes various uncertainties in the assumed dust temperature, opacity, and measured flux.
Among these sources, MM1 is found as a main core associated with the dense gas (see Section~\ref{sec:line}). 

\subsection{Molecular Line Emission}
\label{sec:line}
\begin{figure}
\centering
\includegraphics[width=7.5cm]{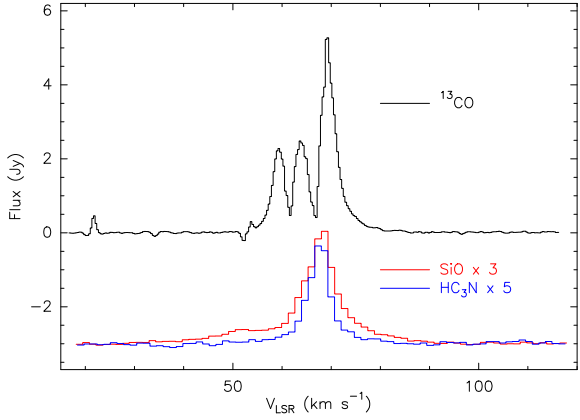}
\caption{ALMA $^{13}$CO, SiO and HC$_{3}$N flux spectra integrated over the area covering MM1 and the outflow. The SiO and HC$_{3}$N spectra are shifted along the ordinate axis for clarity.} 
\label{fig:spectra}
\end{figure}

%
%In general, high density gas tracer HC$_{3}$N depicts dense core(s) and the shock tracer SiO emission is examined to study the outflow(s) and jet(s). 
Figure~\ref{fig:spectra} presents the profiles of the {$^{13}$CO}, HC$_{3}$N and SiO emission toward MM1, indicating the existence of extended non-gaussian wings {(indicative of high-velocity outflows) in all of these lines}. 

\begin{figure}
\centering
\includegraphics[width=8.5cm]{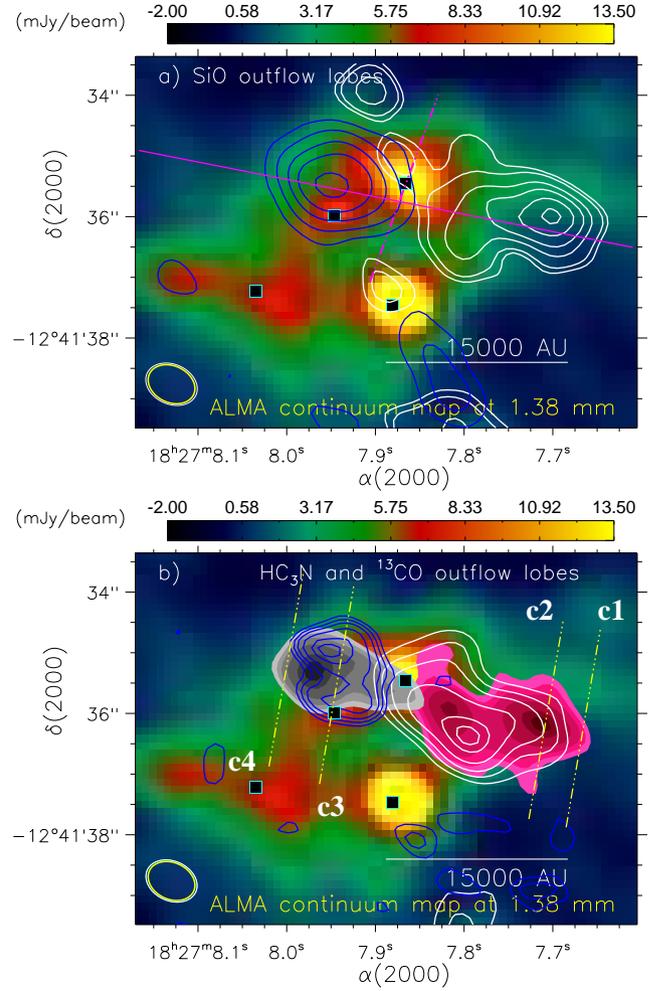}
\caption{{Overlay of the outflow lobes of the molecular emission on the ALMA 1.38 mm continuum image.  
a) The observed outflow in the SiO emission. The contours of the red-shifted component at [74, 94] km s$^{-1}$ (in white) are 
at (0.18, 0.24, 0.4, 0.55, 0.75, 0.95) $\times$ 338.8 mJy beam$^{-1}$ km s$^{-1}$. 
The contours of the blue-shifted component at [28, 61] km s$^{-1}$ (in blue) are at (0.1, 0.22, 0.4, 0.6, 0.9) $\times$ 906.2 mJy beam$^{-1}$ km s$^{-1}$. 
PV diagrams along the dotted-dashed line and the solid line are presented in Figures~\ref{fig2} and~\ref{fig4}, respectively. 
b) The observed outflows in the $^{13}$CO (solid curves) 
and HC$_{3}$N (filled contours) emission. 
The $^{13}$CO emission contours are at (0.3, 0.4, 0.5, 0.65, 0.85, 0.95) $\times$ peak value (i.e., 
438.5 mJy beam$^{-1}$ km s$^{-1}$ for red-shifted component at [74, 90] km s$^{-1}$ (in white) and 64.7 mJy beam$^{-1}$ km s$^{-1}$ for blue-shifted component at [44, 50] km s$^{-1}$ (in blue)). 
Filled contours of the HC$_{3}$N emission are at (0.3, 0.4, 0.5, 0.65, 0.85, 0.95) $\times$ peak value (i.e., 102.4 mJy beam$^{-1}$ km s$^{-1}$ for red-shifted component at [73, 85] km s$^{-1}$ (in pink) and 224.5 mJy beam$^{-1}$ km s$^{-1}$ for blue-shifted component at [40, 63] km s$^{-1}$ (in gray)). PV diagrams along four cuts c1--c4 (see dotted-dashed lines) are presented in Figure~\ref{fig5}. In each panel, ellipses show the synthesized beams (lower left corner).} {The filled squares in both panels indicate the center locations of the continuum sources as shown in Figure~\ref{fig1}b.}
}
\label{fig3}
\end{figure}

%Based on the inspection of the SiO and HC$_{3}$N profiles, we find the red-shifted component at [72.7, 84.8] km s$^{-1}$ and the blue-shifted one at [43.2, 63.3] km s$^{-1}$. 
{In Figures~\ref{fig3}a and~\ref{fig3}b, we show the integrated intensity emission contours of the red-shifted and blue-shifted components in the SiO, $^{13}$CO and HC$_{3}$N emission superimposed on the ALMA 1.38~mm continuum map. For the blue-shifted $^{13}$CO emission we selected the velocity range {in the far wing of the line ([44, 50]~\kms)} free from the additional emission peaks, which are probably produced by some other components in this complex.
All these molecular outflows (extent $\sim$28000~AU {in SiO}) are centered at the continuum source MM1. At the same time the morphology of the emission in different lines is somewhat different. In particular, the $^{13}$CO emission apparently surrounds the HC$_{3}$N emission, especially in the blue lobe.} It is worth mentioning that the outflows seem to be bent.

%

\begin{figure}
\centering
\includegraphics[width=7.5cm]{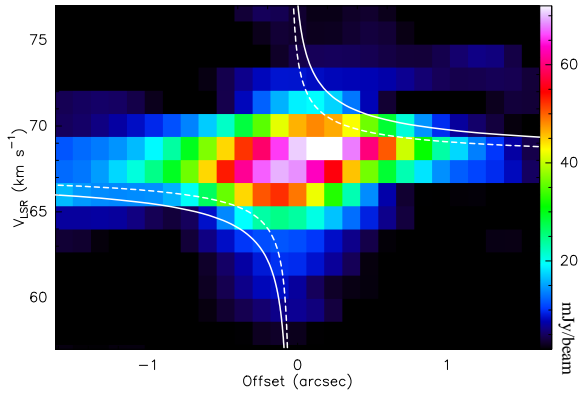}
\caption{PV diagram in the HC$_{3}$N line at the PA of 160$\degr$ across ``MM1" {(perpendicular to the outflow; see a dotted-dashed line in Figure~\ref{fig3}a).} The curves display the Keplerian rotation around the central mass of $M\sin^{2}i$ = 8 M$_{\odot}$ (dashed) and $M\sin^{2}i$ = 12 M$_{\odot}$ (solid).} 
\label{fig2}
\end{figure}

%Through the HC$_{3}$N profile, we find that the line core HC$_{3}$N emission can be traced in a velocity range of [63.3, 72.7] km s$^{-1}$. 
%The integrated intensity map of the HC$_{3}$N emission is displayed in Figure~\ref{fig2}a, where ellipses highlight the positions of the continuum sources. The line core HC$_{3}$N emission is well depicted toward the continuum source MM1.
Figure~\ref{fig2} displays the position-velocity (PV) diagram in the HC$_{3}$N line at the position angle of 160$\degr$ across MM1 (perpendicular to the outflow), allowing to explore the kinematics of the continuum source. 
{This PV diagram is consistent with a Keplerian-like rotation (although the red-shifted part of the emission is practically missing), which may refer to the probable disk in MM1 (unresolved in the present data).
%On the basis of the rotation velocity information, 
The dynamical central mass of the core is determined to be $M \sim 8/\sin^{2}i$~M$_{\odot}$, where $i$ is the unknown disk inclination (with a high uncertainty due to the insufficient angular resolution).} 

\begin{figure}
\centering
\includegraphics[width=8.0cm]{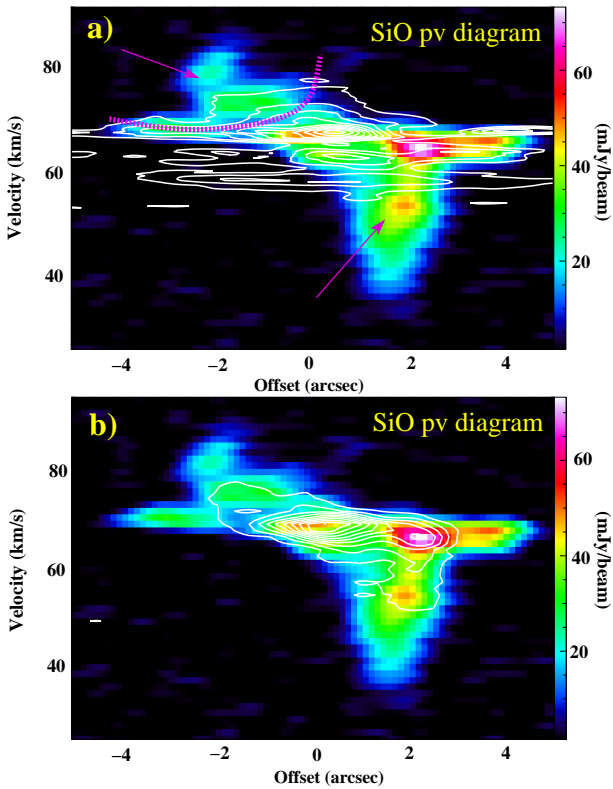}
\caption{a) PV diagram in the SiO line at the PA of 81$\degr$ across ``MM1" {(see a solid line in 
Figure~\ref{fig3}a).} The contours show the $^{13}$CO emission. {The contour levels are at (0.05, 0.15, ..., 0.95) $\times$ 456~mJy\,beam$^{-1}$.} The arrows indicate the jet-like outflow and the dashed curve indicates the probable wide-angle wind. 
b) Overlay of HC$_{3}$N PV contours on the PV diagram in the SiO line. {The contour levels are at (0.1, 0.2, ..., 0.9) $\times$ 69~mJy\,beam$^{-1}$.}}
\label{fig4}
\end{figure}

Figure~\ref{fig4}a displays the PV diagram in the SiO line along the outflow (at the position angle of 81$\degr$) across MM1.
The PV diagram is also overlaid with the $^{13}$CO PV contours (in white). 
In Figure~\ref{fig4}b, we present the overlay of the HC$_{3}$N PV contours on the PV diagram in the SiO line. 
{The PV diagrams indicate the existence of two components:}
a fast jet-like one (with a large extent in velocity) and a slower, more extended spatially component (see arrows and a dashed curve in Fig.~\ref{fig4}a). 
The $^{13}$CO data {may be} consistent with a wide-angle wind picture for the slow component, which shows a structure similar to that expected in this case \citep[for comparison see Figure~2 in][]{arce07}. 
The fast SiO outflow component coincides with the HC$_{3}$N outflow. The PV diagram for this component is typical for the jet-driven bow shock model \citep{arce07}.

\begin{figure}
\includegraphics[width=\columnwidth]{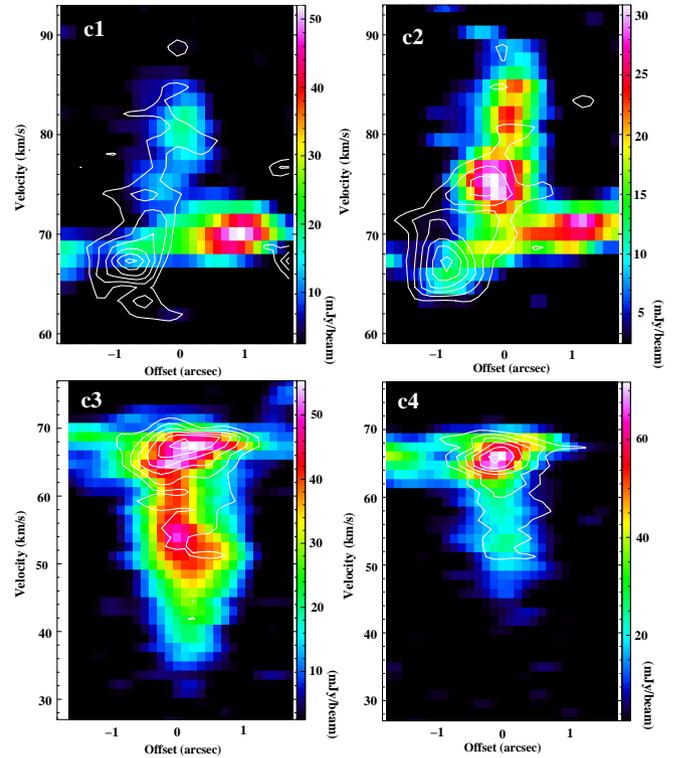}
\caption{PV diagrams in the SiO line correspond to four cuts (i.e., c1--c4) across the outflow lobes 
(see dotted-dashed lines in Figure~\ref{fig3}b). The contours (in white) show the HC$_{3}$N emission. {The contour levels are at (0.2, 0.35, 0.5, 0.65, 0.8, 0.95) $\times$ peak value (15, 28, 36, 29~mJy\,beam$^{-1}$ for the cuts c1, c2, c3 and c4, respectively.)}
}
\label{fig5}
\end{figure}

Figure~\ref{fig5} shows the PV diagrams in the SiO line at four cuts across the outflow lobes (i.e., c1--c4 in Figure~\ref{fig3}b), allowing us to examine the transverse structure of the outflow. 
{The cuts c1 and c2 are selected toward the red-shifted lobe, while the cuts c3 and c4 are chosen in the direction of the blue-shifted lobe. Each PV diagram is also overlaid with the HC$_{3}$N PV contours. 
These PV diagrams show a narrow fast and a wide slow components in both lobes. One can see velocity gradients across the outflow lobes in both components. These gradients are especially clear in the red-shifted outflow lobe (see panels ``c1" and ``c2"). It has the same sign as in the core. In the blue-shifted fast lobe the gradient is either absent or slightly opposite (see panels ``c3" and ``c4").}

\section{Discussion and Conclusion}
\label{sec:discussion}
Outflows/jets are often explained by the X-wind model {and disk-wind model \citep[e.g.,][]{arce07,frank14}.} Previously, using the water maser observations, \citet{torrelles11} reported a two-wind outflow (i.e., a fast narrow jet and a slower wide-angle outflow) from a massive protostar Cep A HW2. A similar picture is observed in S255IR-SMA1 harboring a 20~M$_\odot$ MYSO \citep{zinchenko20}. The observed two flavours of ejection from MYSOs are crucial inputs for unifying the jet-driven and wind-driven scenarios of molecular outflows in MYSOs, which is one of the open research topics in star formation. However, such event is rare in the literature. Hence, the most striking outcome of this work is the simultaneous detection of a wide-angle wind and a narrow jet-like outflow, which are driven by the MYSO G18.88MME embedded in the continuum core MM1 (mass $\sim${13--18} M$_{\odot}$). 
%We estimate the dynamical central mass of MM1 at $\ga$8 M$_{\odot}$, which is higher or comparable with the stellar mass of G18.88MME (i.e., $\sim$8 M$_{\odot}$). This may hint at a very massive disk around the MYSO. However, the uncertainties of these estimates are too high for a firm conclusion. Anyway, the closeness of the stellar mass and dynamical mass estimates indicates that the inclination angle of the suggested disk is probably large.
%These results suggest the onset of the accretion process in G18.88MME. 
%The resolution of the adopted ALMA data is insufficient to resolve the accretion disk around the MYSO.

The mass of the red-shifted outflow lobe estimated from the $^{13}$CO spectrum is $\sim 1\times (T_\mathrm{ex}/100\,\mathrm{K})$~M$_{\odot}$ assuming a normal $^{13}$CO abundance (we have no estimate of the $^{13}$CO excitation temperature $ T_\mathrm{ex} $). The blue part of the $^{13}$CO spectrum is contaminated by the {emission peaks which prevent such an estimate}. The terminal line-of-sight velocity of the outflow as observed in the SiO line is $ \sim 25 $~\kms\ in the red lobe and $ \sim 40 $~\kms\ in the blue lobe. The SiO relative abundance estimated {from the ratio of the SiO and $^{13}$CO line wing intensities toward the SiO emission peaks} is $ \sim 5\times 10^{-9} $ in the red lobe and $ \sim 10^{-8} $ in the blue lobe under the assumption of {low optical depth and} equal LTE excitation of SiO and $^{13}$CO {with $T_\mathrm{ex}>E_\mathrm{u}/k$}. These values are among the highest SiO abundances observed in the outflows in HMSF regions \citep{zinchenko20}.

In general, the HC$_{3}$N(24--23) line is known as a very good tracer of warm and dense gas in star-forming regions \citep{lis93}. The critical density of this transition is $\sim 3\times 10^6$~cm$^{-3}$ using the collisional rates obtained by \citet{Faure16}. Its detection in outflows is very rare and indicates a very high density in the fast jet-like outflow. For the HC$_{3}$N abundance we obtain the values of $ \sim 10^{-9} $ in the red lobe and {a factor of 2} higher in the blue lobe (under the same assumptions as for SiO).
%One requires a very high molecular column density for detection of the HC$_{3}$N(24--23) line. 
%Hence, the detection of the HC$_{3}$N outflow is an extremely rare and remarkable event in our target MYSO \citep[e.g.,][]{lis93}.
%Furthermore, the HC$_{3}$N outflow spatially matches with the slower wide-angle SiO outflow in MM1. 

{Our outcomes also show velocity gradients across the outflow lobes (see Figure~\ref{fig5})}, which can be interpreted as rotation. It is worth noting that such interpretation is not unique. Some asymmetries can produce similar velocity gradients \citep[e.g.,][]{DeColle16}. The picture is complicated by the fact that our outflow is apparently bent. Such bending may hint at the disk precession, which can be a consequence of a binary nature of this system {\citep{Monin07}}. In this case the rotation is combined with a helical motion which can produce a complicated velocity pattern. Perhaps this can explain the difference in the appearance of the red-shifted and blue-shifted outflow lobes with the absence or even an opposite sign of the velocity gradient in the fast component of the latter one.

If we interpret the velocity gradient in the red-shifted outflow lobe as a rotation, {we can {try to} estimate the launching radius of the outflow} following the approach suggested in \citet{Anderson03}. {At the cut ``c2" (Fig.~\ref{fig5}) the total velocity span is about 20~\kms\ at the offset interval of about 0.4~arcsec, which corresponds to 2000~AU. This implies the specific angular momentum of $\sim 10000$~AU\,\kms.} This is a very high value, much higher than observed in nearby low-mass objects \citep[e.g.,][]{Zhang18}. 
{According to this model the launching radius is about $30\times(v_\mathrm{p}/100$~\kms)$^{-4/3}$~AU, where $v_\mathrm{p}$ is the poloidal velocity. For the typical values of $v_\mathrm{p}$ for jets from $\sim$100~\kms\ to $\sim$1000~\kms\ \citep[e.g.][]{Anglada18} the launching radius varies from $\sim$30~AU to $\sim$1.4~AU.}
%Assuming the poloidal velocity of $v_\mathrm{p} \sim 100$~\kms, which is a rather typical value, we obtain the launching radius of about 30~AU. This estimate depends on $v_\mathrm{p}$ as $v_\mathrm{p}^{-4/3}$. %i.e. for the poloidal velocity 2 times higher the launching radius will be 2.5 times smaller. 
%For $v_\mathrm{p} \sim 500$~\kms, which is close to the highest observed velocities in the jets, the launching radius is about 3.5~AU. 
These estimates indicate the disk wind as a launching mechanism for the outflow. {However, they should be considered as very preliminary since the data do not permit us to see the morphology of the outflow lobes and to judge whether the observed high-velocity gas is ejected from the disk or represents the entrained material. So fast rotation on such large scales needs a confirmation.}

%Such a combination of observed findings in G18.88MME makes it a special MYSO in the earliest phase of evolution.

There is a question whether {the two} components have separate origins or perhaps we see a transformation of the fast outflow into a slower one. A larger extent of the slow component is an argument against the {latter suggestion}. In principle there can be episodic ejections but the data do not provide any support for such suggestion. The jet-driven bow shock model for the fast outflow implies an existence of the underlying ionized jet. It can be revealed by high resolution radio and/or IR observations.
%The dependence of the velocity on distance for the fast component is more or less typical for a jet bow shock \citep{arce07} but an interesting feature is its non-monotonic behavior at the largest distances. It can be related to a braking of the fast outflow. Then, it is not excluded that it can be explained by the projection effects. 

Finally, we can conclude that G18.88MME represents a unique case of a massive YSO driving a very dense probably rotating fast jet-like molecular outflow surrounded by a slower wide-angle wind. {This object deserves an investigation at a much higher angular resolution.}

%Both of the components appear to be rotating (see Figure~\ref{fig5}). 
%It is not clear whether the SiO emission at 70 km s$^{-1}$ belongs to the outflow or just to the ambient medium, which cannot be excluded taking into account a complicated structure of the region. It is very prominent at the end of the red lobe but has no counterpart in HC$_{3}$N. Furthermore, there is a signature of the precession of the jet. 
%The main puzzle is the opposite rotation (if it is a rotation) of the blue-shifted high-velocity component. Actually the outflow looks somewhat bent which is usually interpreted as a sign of the disk precession. A precession in turn may be a result of a binary nature of the system. In this case we deal with a helical motion which may influence the observed velocity gradients.
%

\subsection*{Data availability}
The ALMA continuum data underlying this article are available from the publicly accessible JVO ALMA FITS archive\footnote[2]{http://jvo.nao.ac.jp/portal/alma/archive.do/}.
The {\it Spitzer} 24 $\mu$m continuum map underlying this article is available from the publicly accessible NASA/IPAC infrared science archive\footnote[3]{https://irsa.ipac.caltech.edu/frontpage/}.
The published IRAM 1.2 mm continuum map underlying this article is available from the website of the VizieR Service\footnote[4]{https://vizier.u-strasbg.fr/viz-bin/VizieR}.

\section*{Acknowledgments}
{We are very grateful to the anonymous referee for the helpful comments.}
I.I.Z. acknowledges the support by the Russian Science Foundation (grant No. 17-12-01256). 
The research work at Physical Research Laboratory is funded by the Department of Space, Government of India. 
%TB is supported by the Satyendra Nath Bose National Centre for Basic Sciences, Kolkata, India, under the Department of Science and Technology (DST), Government of India. 
DKO acknowledges the support of the Department of Atomic Energy, Government of India, under project Identification No. RTI 4002.
This work is based [in part] on observations made with the {\it Spitzer} Space Telescope, which is operated by the Jet Propulsion Laboratory, California Institute of Technology under a contract with NASA. 
This paper makes use of the following archival ALMA data: ADS/JAO.ALMA\#2017.1.00983.S. ALMA is a partnership of ESO (representing its member states), NSF (USA) and NINS (Japan), together with NRC (Canada), MOST and ASIAA (Taiwan), and KASI (Republic of Korea), in cooperation with the Republic of Chile. The Joint ALMA Observatory is operated by ESO, AUI/NRAO and NAOJ. 
%In addition, publications from NA authors must include the standard NRAO acknowledgement: The National Radio Astronomy Observatory is a facility of the National Science Foundation operated under cooperative agreement by Associated Universities, Inc.
%%



\bibliographystyle{mnras}
\bibliography{sdc18} % if your bibtex file is called example.bib

%\begin{thebibliography}{99}
%\end{thebibliography}

%%%%%%%%%%%%%%%%%%%%%%%%%%%%%%%%%%%%%%%%%%%%%%%%%%

%%%%%%%%%%%%%%%%% APPENDICES %%%%%%%%%%%%%%%%%%%%%

%\appendix



%%%%%%%%%%%%%%%%%%%%%%%%%%%%%%%%%%%%%%%%%%%%%%%%%%

% Don't change these lines
\bsp	% typesetting comment
\label{lastpage}
\end{document}